\documentclass[conference,a4paper]{IEEEtran}
\usepackage{xcolor}
\usepackage{balance}

\ifCLASSINFOpdf
  \usepackage[pdftex]{graphicx}
\else
  \usepackage[dvips]{graphicx}
\fi
\usepackage{amsmath,amssymb}
\ifCLASSOPTIONcompsoc
 \usepackage[caption=false,font=normalsize,labelfont=sf,textfont=sf]{subfig}
\else
 \usepackage[caption=false,font=footnotesize]{subfig}
\fi

\usepackage{bm}

\newcommand{\eg}{\textit{e.g.}\/, }
\newcommand{\cf}{\textit{cf.}\/, }

\providecommand*{\mrm}[1]{\mathrm{#1}}
\providecommand*{\unit}[1]{\ensuremath{\mrm{\,#1}}}
\providecommand*{\eu}{\ensuremath{\mrm{e}}}
\providecommand*{\iu}{\ensuremath{\mrm{i}}}

\providecommand*{\diff}{\operatorname{d}\!}
\renewcommand{\Re}{\ensuremath{\mrm{Re}}}	
\renewcommand{\Im}{\ensuremath{\mrm{Im}}}	

\newcommand{\maximize}{\mrm{maximize}}

\newcommand{\subto}{\mrm{subject\ to}}

\newcommand{\Ohmps}{\unit{\Omega}/\square}

\begin{document}
%
\title{On the optical theorem and optimal extinction, scattering and absorption in lossy media}

\author{\IEEEauthorblockN{
Sven~Nordebo\IEEEauthorrefmark{1},   
Mats~Gustafsson\IEEEauthorrefmark{2},   
Yevhen~Ivanenko\IEEEauthorrefmark{1}    
}                                     
\IEEEauthorblockA{\IEEEauthorrefmark{1}
Department of Physics and Electrical Engineering, Linn\ae us University,   351 95 V\"{a}xj\"{o}, Sweden. E-mail: sven.nordebo@lnu.se} 
\IEEEauthorblockA{\IEEEauthorrefmark{2}
Department of Electrical and Information Technology, Lund University, Box 118, 221 00 Lund, Sweden.} 
}



\maketitle

\begin{abstract}

This paper reformulates and extends some recent analytical results concerning a new optical theorem
and the associated physical bounds on absorption in lossy media. 
The analysis is valid for any linear scatterer (such as an antenna), consisting of arbitrary materials (bianisotropic, etc.)  and arbitrary geometries,  as long as the scatterer is
circumscribed by a spherical volume embedded in a lossy background medium.
The corresponding formulas are here reformulated and extended to encompass 
magnetic as well as dielectric background media.  
Explicit derivations, formulas and discussions are also given for the corresponding bounds on scattering and extinction.
A numerical example concerning the optimal microwave absorption and scattering in atmospheric oxygen in the 60 GHz communication band
is included to illustrate the theory.


\end{abstract}

\vskip0.5\baselineskip

%

\section{Introduction}

Perhaps counter-intuitive, the presence of external losses in the surrounding medium implies major difficulties
in the formulation of an optical theorem for a general scattering object, and a reasonably simple analytical solution is typically available
only for spheres \cite{Bohren+Gilra1979,Lebedev+etal1999,Sudiarta+Chylek2001,Durant+etal2007a}, see also \cite{Nordebo+etal2019a,Nordebo+etal2019b,Ivanenko+etal2019c}.
Analytical solutions based on the spherical vector wave expansion (Mie theory) \cite{Sudiarta+Chylek2001} have recently been
used to formulate an optical theorem and to derive explicit formulas for the optimal multipole absorption, scattering and extinction of a spherical object embedded
in a lossy medium \cite{Nordebo+etal2019a,Nordebo+etal2019b}.
In  \cite{Ivanenko+etal2019c} has been shown that the theory developed in \cite{Nordebo+etal2019a} is valid not only for the individual multipole absorption
of a rotationally invariant sphere, but also for the absorption of an arbitrary scatterer (consisting of arbitrary materials and structures) embedded inside a spherical inclusion,
and furthermore that the bound is valid also as a summation over all multipole fields (the total field) as long as the losses in the surrounding medium are not vanishing.
In this contribution, we reformulate and extend the analytical results given in \cite{Ivanenko+etal2019c} to encompass 
magnetic as well as dielectric background media, and we give explicit derivations and formulas also for the corresponding bounds on scattering and extinction.
In essence, the theory gives new fundamental bounds on absorption, scattering and extinction that limits any single frequency super resolution effects
for a scatterer in a lossy surrounding medium, with possible applications \eg
with antennas and resonators in communications \cite{Park+Rappaport2007,Hawkins+etal1985,Wang+etal2012,Karlsson2004,Skrivervik2019,Gustafsson+Miloslav2019}, 
plasmonics and metamaterials \cite{Maier2007,Maslovski+etal2016,Valagiannopoulos+Tretyakov2016} 
and medicine \cite{Huang+etal2008,Sassaroli+etal2012,Nordebo+etal2017a}.

\section{Optical theorem for a spherical inclusion in lossy media}
\subsection{Notation and conventions}
The electric and magnetic field intensities $\bm{E}$ and $\bm{H}$
are given in SI-units \cite{Jackson1999} and the time convention for time harmonic fields (phasors) is given by $\eu^{-\iu\omega t}$,
where $\omega$ is the angular frequency and $t$ the time. 
Let $\mu_0$, $\epsilon_0$, $\eta_0$ and $\mrm{c}_0$ denote the permeability, the permittivity, the wave impedance and
the speed of light in vacuum, respectively, and where $\eta_0=\sqrt{\mu_0/\epsilon_0}$ and $\mrm{c}_0=1/\sqrt{\mu_0\epsilon_0}$.
The wave number of vacuum is given by $k_0=\omega\sqrt{\mu_0\epsilon_0}$, and hence $\omega\mu_0=k_0\eta_0$ and $\omega\epsilon_0=k_0\eta_0^{-1}$. 
The definition of the spherical vector waves \cite{Kristensson2016} and their most important properties employed in this paper can be found in \cite[A(1)--A(32)]{Nordebo+etal2019a}.
Here, $l=1,2,\ldots,$ denotes the multipole order, $m=-l,\ldots,l$, the azimuthal index and 
$\tau=1$ indicates a transverse electric (\textrm{TE}) magnetic multipole and $\tau=2$ a transverse magnetic (\textrm{TM}) electric multipole, respectively.
The regular spherical Bessel functions, the Neumann functions, the spherical Hankel functions of the first kind 
and the corresponding Riccati-Bessel functions \cite{Kristensson2016} are denoted $\mrm{j}_l(z)$, $\mrm{y}_l(z)$, $\mrm{h}_l^{(1)}(z)=\mrm{j}_l(z)+\iu\mrm{y}_l(z)$,
$\psi_l(z)=z\mrm{j}_l(z)$ and $\xi_l(z)=z\mrm{h}_l^{(1)}(z)$, respectively, all of order $l$.
The real and imaginary parts and the complex conjugate of a complex number $\zeta$ are denoted by $\Re\left\{\zeta\right\}$, $\Im\left\{\zeta\right\}$ and $\zeta^*$, respectively. 

%

\subsection{Absorption, scattering and extinction}

Consider a scattering problem where an absorbing structure (such as a receiving antenna) is embedded in an infinite homogeneous and isotropic  
background medium having relative permeability $\mu$, relative permittivity $\epsilon$, wave impedance $\eta_0\eta$ where
$\eta=\sqrt{\mu/\epsilon}$  and wave number $k=k_0\sqrt{\mu\epsilon}$. 
The absorbing structure is made of an arbitrary material (general bianisotropic etc.) in arbitrary geometry, and is circumscribed by a minimal sphere of radius $a$.
The scattering body $V$ is considered to be the whole circumscribing sphere with surface denoted $\partial V$.
The background medium is passive and in general lossy, with complex-valued parameters satisfying $\Im\{\mu\}\geq 0$, $\Im\{\epsilon\}\geq 0$
and $\Im\{k\}\geq 0$.  

Let the total fields outside $V$ be decomposed in incident and scattered fields as 
$\bm{E}=\bm{E}_\mrm{i}+\bm{E}_\mrm{s}$ and $\bm{H}=\bm{H}_\mrm{i}+\bm{H}_\mrm{s}$, respectively.
Following \cite{Bohren+Huffman1983,Nordebo+etal2019a,Ivanenko+etal2019c} the optical theorem can now be derived from the power balance
at the surface $\partial V$
\begin{equation}\label{eq:Opticaltheorem}
P_\mrm{a}=-P_\mrm{s}+P_\mrm{t}+P_\mrm{i},
\end{equation}
where $P_\mrm{a}$, $P_\mrm{s}$, $P_\mrm{t}$ and $P_\mrm{i}$ are the absorbed, scattered, extinct (total) and the absorbed incident powers, respectively,
defined by 
\begin{flalign}
P_\mrm{a} & =  -\frac{1}{2}\Re\left\{ \int_{\partial V}\bm{E}\times\bm{H}^* \cdot \hat{\bm{n}}\diff S \right\},  \label{eq:Padef} \\
P_\mrm{s} & =  \frac{1}{2}\Re\left\{ \int_{\partial V}\bm{E}_\mrm{s}\times\bm{H}_\mrm{s}^* \cdot \hat{\bm{n}}\diff S \right\}, \label{eq:Psdef}\\
P_\mrm{t} & =  -\frac{1}{2}\Re\left\{ \int_{\partial V}\left(\bm{E}_\mrm{i}\times\bm{H}_\mrm{s}^* + \bm{E}_\mrm{s}\times\bm{H}_\mrm{i}^*\right)  \cdot \hat{\bm{n}}\diff S \right\}, \label{eq:Ptdef} \\
P_\mrm{i} & =  -\frac{1}{2}\Re\left\{ \int_{\partial V}\bm{E}_\mrm{i}\times\bm{H}_\mrm{i}^* \cdot \hat{\bm{n}}\diff S \right\}, \label{eq:Pidef}
\end{flalign}
and where the surface integrals are defined with an outward unit normal $\hat{\bm{n}}$.

Let $a_{\tau ml}^\mrm{i}$ and $f_{\tau ml}$ denote the multipole coefficients of the incident (regular) and the scattered (outgoing) spherical vector waves, respectively,
as defined in \cite[A(1)]{Nordebo+etal2019a}.
Based on the orthogonality of the spherical vector waves on the spherical surface $\partial V$ 
as given by \cite[A(31) and A(32)]{Nordebo+etal2019a}, it can be shown that 
\begin{flalign}
P_\mrm{s} & =  \frac{\Re\{1/\eta\}}{2\left|k\right|^2\eta_0}\sum_{\tau,m,l}A_{\tau l}\left|f_{\tau ml}\right|^2,  \label{eq:Psexprout}\\
P_\mrm{t} & =  \frac{\Re\{1/\eta\}}{2\left|k\right|^2\eta_0}\sum_{\tau,m,l}2\Re\{B_{\tau l}a_{\tau ml}^\mrm{i*}f_{\tau ml}\}, \label{eq:Ptexprout} \\
P_\mrm{i} & =  \frac{\Re\{1/\eta\}}{2\left|k\right|^2\eta_0}\sum_{\tau,m,l}C_{\tau l}\left|a_{\tau ml}^\mrm{i}\right|^2, \label{eq:Piexprout}
\end{flalign}
where
\begin{flalign}
A_{\tau l} & =  \frac{1}{\Re\{1/\eta\}}\left\{
\begin{array}{ll}
-\Im\{\xi_l\xi_l^{\prime *}/\eta^*\} & \tau=1, \vspace{0.2cm} \\
\Im\{\xi_l^{\prime}\xi_l^{*}/\eta^*\} & \tau=2,
\end{array}
\right. \label{eq:Atauldef}
\\
B_{\tau l} & =  \frac{1}{2\iu \Re\{1/\eta\}}\left\{
\begin{array}{ll}
\xi_l\psi_l^{\prime *}/\eta^*-\psi_l^*\xi_l^\prime/\eta & \tau=1, \vspace{0.2cm} \\
-\xi_l^\prime\psi_l^{*}/\eta^*+\psi_l^{\prime *}\xi_l/\eta & \tau=2,
\end{array}
\right.\label{eq:Btauldef}
\\
C_{\tau l} & =  \frac{1}{\Re\{1/\eta\}}\left\{
\begin{array}{ll}
\Im\{\psi_l\psi_l^{\prime *}/\eta^*\} & \tau=1, \vspace{0.2cm} \\
-\Im\{\psi_l^{\prime}\psi_l^{*}/\eta^*\} & \tau=2,
\end{array}
\right.\label{eq:Ctauldef}
\end{flalign}
for $\tau=1,2$ and $l=1,\ldots,\infty$, and where the arguments of the Riccati-Bessel functions are $z=ka$.
Notice the modification in the definition of these parameters in terms of the relative wave impedance $\eta$ (due to the presence of the magnetic property), in comparison to
\cite[Eqs.~(9) through (11)]{Nordebo+etal2019a} and \cite[Eqs.~(37) through (42)]{Ivanenko+etal2019c}.
Similar as in \cite{Nordebo+etal2019a,Ivanenko+etal2019c}, it can be readily shown that $A_{\tau l}> 0$, $C_{\tau l}\geq 0$
and it is noticed that $B_{\tau l}$ is in general a complex-valued constant. 
As expected, it is possible to show that \eqref{eq:Atauldef} through \eqref{eq:Ctauldef} are invariant to the change $\mu\leftrightarrow\epsilon$
under the summation over $\tau=1,2$.
For a lossless medium ($k$ and $\eta$ are real-valued), it can furthermore be shown
that $A_{\tau l}=1$, $B_{\tau l}=-1/2$ and $C_{\tau l}=0$, see \cite[p.~3]{Nordebo+etal2019a}, in agreement with \eg \cite[Eq.~(7.18)]{Kristensson2016}.

\section{Optimal cross sections}

\subsection{Optimal absorption}
The optimal absorption presented here has previously been derived in \cite[Eqs.~(43) through (51)]{Ivanenko+etal2019c}.
The approach is briefly summarized below to provide a framework for a more general setting including scattering and extinction, as well as to incorporate
both electric and magnetic surrounding media. 

Consider the power absorbed from a single partial wave with fixed multi-index $n=(\tau, m,l)$,
\begin{multline}\label{eq:Papartwave}
P_{\mrm{a},n}=\frac{\Re\{1/\eta\}}{2\left|k\right|^2\eta_0}\left(-A_{\tau l}\left|f_{n}\right|^2 
 +2\Re\{B_{\tau l}a_{n}^\mrm{i*}f_{n}\} \right. \\
 \left. +C_{\tau l}\left|a_{n}^\mrm{i}\right|^2\right),
\end{multline}
where we have employed the optical theorem \eqref{eq:Opticaltheorem} as well as \eqref{eq:Psexprout} through \eqref{eq:Piexprout}.
Let the scattering coefficients $f_{n}$ be given by the T-matrix \cite[Eq.~(7.34)]{Kristensson2016} for an arbitrary linear scatterer inside the spherical surface $\partial V$, so that
\begin{equation}\label{eq:T-matrixdef}
f_{n}=\sum_{n^\prime}
T_{n,n^\prime}a_{n^\prime}^\mrm{i}.
\end{equation}
It is observed that \eqref{eq:Papartwave} is a concave function of the complex-valued T-matrix elements $T_{n,n^\prime}$ with respect to the primed index $n^\prime$.
Hence, we can employ the following complex derivative (for fixed $n$)
\begin{equation}\label{eq:cplxderivative}
\frac{\partial }{\partial T_{n,n^\prime}}=a_{n^\prime}^\mrm{i}\frac{\partial }{\partial f_{n}},
\end{equation}
to find the stationarity condition 
\begin{equation}\label{eq:stationarity}
a_{n^{\prime}}^\mrm{i}
\sum_{n^{\prime\prime}}a_{n^{\prime\prime}}^{\mrm{i}*}T_{n,n^{\prime\prime}}^*
=\frac{B_{\tau l}}{A_{\tau l}}a_{n}^{\mrm{i}*}a_{n^{\prime}}^\mrm{i},
\end{equation}
which is an infinite-dimensional linear system of equations in $T_{n,n^{\prime\prime}}$ for a given $n$, see also \cite[Eqs.~(43) through (45)]{Ivanenko+etal2019c}. 
For a truncated (finite-dimensional) partial wave expansion the unique minimum norm (pseudo-inverse) solution to \eqref{eq:stationarity} is given by
\begin{equation}\label{eq:Tmatsol}
T_{n,n^\prime}^{\rm opt}=
\frac{B_{\tau l}^*}{A_{\tau l}\|\bm{a}^\mrm{i}\|^2}a_{n}^{\mrm{i}}a_{n^{\prime}}^{\mrm{i}*},
\end{equation}
where $\|\bm{a}^\mrm{i}\|^2=\sum_{n}\left| a_{n}^\mrm{i}\right|^2$.
It is noted that the unique pseudo-inverse in \eqref{eq:Tmatsol} can be extended by adding a sequence from the corresponding null space as
\begin{equation}\label{eq:nullspace}
T_{n,n^\prime}^{\rm opt}\rightarrow T_{n,n^\prime}^{\rm opt}+t_{n^\prime} \quad \textrm{where} \quad \sum_{n^\prime}a_{n^{\prime}}^{\mrm{i}}t_{n^\prime}=0,
\end{equation}
and which hence satisfies the stationarity condition \eqref{eq:stationarity} and corresponds to the same scattering coefficient $f_n$ as in \eqref{eq:T-matrixdef}.

By completing the squares based on \eqref{eq:Papartwave}, \eqref{eq:T-matrixdef} and \eqref{eq:Tmatsol}, it can be shown that
\begin{multline}\label{eq:completesquares}
P_{\mrm{a},n}=\frac{\Re\{1/\eta\}}{2\left|k\right|^2\eta_0}\Bigg(    
-A_{\tau l}\left| \sum_{n^\prime}\left(T_{n,n^\prime}-
\frac{B_{\tau l}^*}{A_{\tau l}\|\bm{a}^\mrm{i}\|^2}a_{n}^\mrm{i}a_{n^\prime}^{\mrm{i}*}\right)
a_{n^\prime}^\mrm{i}  \right|^2 \\
\left. +\left(\frac{\left| B_{\tau l} \right|^2}{A_{\tau l}}+C_{\tau l}\right)\left| a_{n}^\mrm{i}\right|^2\right).
\end{multline}
From the concavity of this expression ($A_{\tau l}>0$), it follows that the last term gives the optimal absorption
\begin{equation}
P_{\mrm{a},n}^\mrm{opt}=
\frac{\Re\{1/\eta\}}{2\left|k\right|^2\eta_0}  
\left(\frac{\left| B_{\tau l} \right|^2}{A_{\tau l}}+C_{\tau l}\right)\left| a_{n}^\mrm{i}\right|^2.
\end{equation}

Consider now a plane wave $\bm{E}_\mrm{i}(\bm r)=\bm{E}_0\eu^{\mrm{i} k\hat{\bm{k}}\cdot\bm{r}}$ 
with vector amplitude $\bm{E}_0$, propagation direction $\hat{\bm{k}}$ and power intensity (at $\bm{r}=\bm{0}$)
$I_\mrm{i}=\left| \bm{E}_0\right|^2\Re\{1/2\eta_0\eta\}$.
The corresponding multipole expansion coefficients are given by
$a_{\tau ml}^\mrm{i}=4\pi\iu^{l-\tau+1}\bm{E}_0\cdot{\bf A}_{\tau ml}^*(\hat{\bm{k}})$, 
where ${\bf A}_{\tau ml}(\hat{\bm{k}})$ are the vector spherical harmonics, see \cite[Eqs.~(5) and (A4)]{Nordebo+etal2019a}. 
It can also be shown that
\begin{equation}\label{eq:sumoverm}
\sum_{m=-l}^l \left| a_{\tau ml}^\mrm{i} \right|^2=2\pi (2l+1)\left| \bm{E}_0\right|^2,
\end{equation}
for $\tau=1,2$, see \eg \cite[Eqs.~(6)]{Nordebo+etal2019a}.
The optimal normalized absorption cross section is now obtained as
\begin{multline}\label{eq:newboundQa}
Q_\mrm{a}^\mrm{opt}=\frac{1}{\pi a^2 I_\mrm{i}} \sum_n P_{\mrm{a},n}^\mrm{opt}\\
=\frac{2}{\left|ka\right|^2}\sum_{\tau=1}^{2}\sum_{l=1}^\infty(2l+1)
\left(\frac{\left| B_{\tau l} \right|^2}{A_{\tau l}}+C_{\tau l}\right).
\end{multline}
In \cite[Eqs.~(53) through (57)]{Ivanenko+etal2019c} is shown that the coefficients $A_{\tau l}$, $B_{\tau l}$ and $C_{\tau l}$ have factorial increase, are bounded and
have factorial decrease, respectively, and hence that the expansion  in \eqref{eq:newboundQa} converges whenever there are nonzero losses
in the background medium with $\Im\{k\}> 0$. In the lossless case (with $A_{\tau l}=1$, $B_{\tau l}=-1/2$ and $C_{\tau l}=0$) we obtain the divergent series 
\begin{equation}\label{eq:newboundQaLL}
Q_{\mrm{a},L}^\mrm{opt}=\frac{1}{2\left(ka\right)^2}\sum_{\tau=1}^{2}\sum_{l=1}^L(2l+1)=\frac{1}{\left(ka\right)^2}L(L+2),
\end{equation}
indicating the optimal absorption for each multipole order $l$, see also \cite[Eqs.~(19) and (20)]{Nordebo+etal2019a}.

\subsection{Optimal scattering}
The scattered power of a partial wave with fixed multi-index $n=(\tau, m,l)$ is given by
\begin{equation}\label{eq:Psnexpr}
P_{\mrm{s},n}  =  \frac{\Re\{1/\eta\}}{2\left|k\right|^2\eta_0}A_{\tau l}\left|f_{n}\right|^2,
 \end{equation}
 \cf \eqref{eq:Psexprout}.
Consider now the constrained optimization problem
\begin{equation}\label{eq:Psoptprobdef}
\begin{array}{llll}
	& \maximize & & P_{\mrm{s},n} \\    
	& \subto & &  P_{\mrm{a},n}\geq 0,
\end{array}
\end{equation}
where $P_{\mrm{a},n}$ is the corresponding absorbed power given by \eqref{eq:Papartwave}.
This is a problem of maximizing a convex function over a convex set, which can be formulated equivalently as
\begin{equation}\label{eq:Psoptprobdef2}
\begin{array}{llll}
	& \maximize & &  \left|f_{n}\right|^2 \vspace{0.2cm} \\    
	& \subto & &  A_{\tau l}\left|f_{n}\right|^2 -2\Re\{B_{\tau l}a_{n}^\mrm{i*}f_{n}\} \vspace{0.2cm} \\  
	&  & & \hspace{2.5cm} -C_{\tau l}\left|a_{n}^\mrm{i}\right|^2=0,
\end{array}
\end{equation}
where it is a priori known that the constraint must be active. Here, $f_{n}$ is given by \eqref{eq:T-matrixdef}
and the optimization is with respect to the T-matrix elements $T_{n,n^\prime}$ for a fixed multi-index $n$.

The Lagrangian for this problem is
\begin{multline}
L= \left|f_{n}\right|^2+\alpha\left( A_{\tau l}\left|f_{n}\right|^2 -B_{\tau l}a_{n}^\mrm{i*}f_{n}
-B_{\tau l}^*a_{n}^\mrm{i}f_{n}^* \right. \\
\left.  -C_{\tau l}\left|a_{n}^\mrm{i}\right|^2\right),
\end{multline}
where $\alpha$ is the Lagrange multiplier. The condition for stationarity is obtained by applying the complex derivative \eqref{eq:cplxderivative} yielding
\begin{equation}\label{eq:stationarityScatt}
a_{n^{\prime}}^\mrm{i}\sum_{n^{\prime\prime}}a_{n^{\prime\prime}}^{\mrm{i}*}T_{n,n^{\prime\prime}}^*
=\frac{\alpha B_{\tau l}}{1+\alpha A_{\tau l}}a_{n}^{\mrm{i}*}a_{n^{\prime}}^\mrm{i},
\end{equation}
with the unique minimum norm (pseudo-inverse) solution
\begin{equation}\label{eq:TmatsolScatt}
T_{n,n^\prime}^{\rm opt}=
\frac{\beta B_{\tau l}^*}{\|\bm{a}^\mrm{i}\|^2}a_{n}^{\mrm{i}}a_{n^{\prime}}^{\mrm{i}*},
\end{equation}
where 
\begin{equation}\label{eq:betaScatt}
\beta=\frac{\alpha}{1+\alpha A_{\tau l}}.
\end{equation}
The parameter $\beta$ (for fixed $n$) is determined by inserting \eqref{eq:T-matrixdef} into the constraint equation in \eqref{eq:Psoptprobdef2}
and completing the squares based on \eqref{eq:TmatsolScatt}. The following quadratic condition appears after some labor
\begin{multline}\label{eq:completesquaresScatt}
A_{\tau l}\left| \sum_{n^\prime}\left(T_{n,n^\prime}-
\frac{\beta B_{\tau l}^*}{\|\bm{a}^\mrm{i}\|^2}a_{n}^\mrm{i}a_{n^\prime}^{\mrm{i}*}\right)a_{n^\prime}^\mrm{i}  \right|^2 \\
+(A_{\tau l}\beta-1)2\Re\left\{B_{\tau l}a_{n}^{\mrm{i}*}\sum_{n^\prime}T_{n,n^\prime} a_{n^\prime}^\mrm{i} \right\} \\
-\left(\beta^2A_{\tau l} \left|B_{\tau l}\right|^2+C_{\tau l} \right)\left| a_{n}^\mrm{i} \right|^2=0.
\end{multline}
It is seen that the condition \eqref{eq:completesquaresScatt} is invariant to the choice of null space solutions as expressed in \eqref{eq:nullspace}.
Hence, by inserting \eqref{eq:TmatsolScatt} into \eqref{eq:completesquaresScatt} the following quadratic equation is obtained
\begin{equation}\label{eq:betaquadreq}
\beta^2-2\beta\frac{1}{A_{\tau l}}-\frac{C_{\tau l}}{A_{\tau l}\left|B_{\tau l}\right|^2}=0,
\end{equation}
yielding the solution
\begin{equation}\label{eq:betasolScatt}
\beta_{\tau l}=\frac{1}{A_{\tau l}}+\sqrt{\frac{1}{A_{\tau l}^2}+\frac{C_{\tau l}}{A_{\tau l}\left|B_{\tau l}\right|^2}},
\end{equation}
where it is the larger root that yields the maximally scattered power. The corresponding optimal scattering coefficient
is obtained by inserting \eqref{eq:TmatsolScatt} into \eqref{eq:T-matrixdef} yielding
\begin{equation}\label{eq:fnsolScatt}
f_n^\mrm{opt}=\beta_{\tau l} B_{\tau l}^*a_{n}^\mrm{i}.
\end{equation}
The optimal normalized scattering cross section can now be obtained as
\begin{multline}\label{eq:newboundQs}
Q_\mrm{s}^\mrm{opt}=\frac{1}{\pi a^2 I_\mrm{i}} \sum_n P_{\mrm{s},n}^\mrm{opt}\\
=\frac{2}{\left|ka\right|^2}\sum_{\tau=1}^{2}\sum_{l=1}^\infty(2l+1)\beta_{\tau l}^2A_{\tau l}\left| B_{\tau l} \right|^2,
\end{multline}
where \eqref{eq:Psnexpr}, \eqref{eq:fnsolScatt} and \eqref{eq:sumoverm} have been used.
Based on the asymptotic behavior of the coefficients $A_{\tau l}$, $B_{\tau l}$ and $C_{\tau l}$, it is readily seen
that the expansion  in \eqref{eq:newboundQs} converges whenever there are nonzero losses
in the background medium with $\Im\{k\}> 0$, \cf \cite[Eqs.~(53) through (57)]{Ivanenko+etal2019c}. 
In the lossless case (with $A_{\tau l}=1$, $B_{\tau l}=-1/2$ and $C_{\tau l}=0$) we obtain the divergent series 
$Q_{\mrm{s},L}^\mrm{opt}=4L(L+2)/(k_0a)^2$, similar as in \eqref{eq:newboundQaLL}.

\subsection{Optimal extinction}
The optimal extinction can be obtained similarly to the case with optimal scattering as described above.
The extinct (total) power of a partial wave with fixed multi-index $n=(\tau, m,l)$ is given by
\begin{equation}\label{eq:Ptnexpr}
P_{\mrm{t},n}  =  \frac{\Re\{1/\eta\}}{2\left|k\right|^2\eta_0}2\Re\{B_{\tau l}a_{n}^\mrm{i*}f_{n}\},
 \end{equation}
 \cf \eqref{eq:Ptexprout}.
Consider now the constrained optimization problem
\begin{equation}\label{eq:Ptoptprobdef}
\begin{array}{llll}
	& \maximize & & P_{\mrm{t},n} \\    
	& \subto & &  P_{\mrm{a},n}\geq 0,
\end{array}
\end{equation}
where $P_{\mrm{a},n}$ is the corresponding absorbed power given by \eqref{eq:Papartwave}.
This is a problem of maximizing a convex (linear) function over a convex set, which can be formulated equivalently as
\begin{equation}\label{eq:Ptoptprobdef2}
\begin{array}{llll}
	& \maximize & & 2\Re\{B_{\tau l}a_{n}^\mrm{i*}f_{n}\}  \vspace{0.2cm} \\    
	& \subto & &  A_{\tau l}\left|f_{n}\right|^2 -2\Re\{B_{\tau l}a_{n}^\mrm{i*}f_{n}\} \vspace{0.2cm} \\  
	&  & & \hspace{2.5cm} -C_{\tau l}\left|a_{n}^\mrm{i}\right|^2=0,
\end{array}
\end{equation}
and where the constraint is a priori known to be active.
Again, $f_{n}$ is given by \eqref{eq:T-matrixdef} and the optimization is with respect to the T-matrix elements $T_{n,n^\prime}$ for a fixed multi-index $n$.

The Lagrangian for this problem is
\begin{multline}
L= B_{\tau l}a_{n}^\mrm{i*}f_{n}+B_{\tau l}^*a_{n}^\mrm{i}f_{n}^* \\
+\alpha\left( A_{\tau l}\left|f_{n}\right|^2 -B_{\tau l}a_{n}^\mrm{i*}f_{n}
-B_{\tau l}^*a_{n}^\mrm{i}f_{n}^*-C_{\tau l}\left|a_{n}^\mrm{i}\right|^2\right),
\end{multline}
where $\alpha$ is the Lagrange multiplier. The condition for stationarity is obtained by applying the complex derivative \eqref{eq:cplxderivative} yielding
\begin{equation}\label{eq:stationarityExt}
a_{n^{\prime}}^\mrm{i}\sum_{n^{\prime\prime}}a_{n^{\prime\prime}}^{\mrm{i}*}T_{n,n^{\prime\prime}}^*
=\frac{\alpha-1}{\alpha}\frac{B_{\tau l}}{A_{\tau l}}a_{n}^{\mrm{i}*}a_{n^{\prime}}^\mrm{i},
\end{equation}
with the unique minimum norm (pseudo-inverse) solution
\begin{equation}\label{eq:TmatsolExt}
T_{n,n^\prime}^{\rm opt}=
\frac{\beta B_{\tau l}^*}{\|\bm{a}^\mrm{i}\|^2}a_{n}^{\mrm{i}}a_{n^{\prime}}^{\mrm{i}*},
\end{equation}
where 
\begin{equation}\label{eq:betaExt}
\beta=\frac{\alpha-1}{\alpha A_{\tau l}}.
\end{equation}
It is observed that the solution \eqref{eq:TmatsolExt} is identical to \eqref{eq:TmatsolScatt} in terms of the parameter $\beta$,
and it is merely the relation to $\alpha$ that differs, \cf \eqref{eq:betaExt} and \eqref{eq:betaScatt}.
This means that the optimal solutions in terms of $\beta$ and $f_n$ in \eqref{eq:betasolScatt} and \eqref{eq:fnsolScatt} are also valid in the
case of optimal extinction.
The optimal normalized extinction cross section can now be obtained as
\begin{equation}\label{eq:newboundQt}
Q_\mrm{t}^\mrm{opt}=\frac{1}{\pi a^2 I_\mrm{i}} \sum_n P_{\mrm{t},n}^\mrm{opt} 
=\frac{4}{\left|ka\right|^2}\sum_{\tau=1}^{2}\sum_{l=1}^\infty(2l+1)\beta_{\tau l} \left| B_{\tau l} \right|^2,
\end{equation}
where \eqref{eq:Ptnexpr}, \eqref{eq:fnsolScatt} and \eqref{eq:sumoverm} have been used.
Based on the asymptotic behavior of the coefficients $A_{\tau l}$, $B_{\tau l}$ and $C_{\tau l}$, it is readily seen
that the expansion in \eqref{eq:newboundQt} converges whenever there are nonzero losses
in the background medium with $\Im\{k\}> 0$, \cf \cite[Eqs.~(53) through (57)]{Ivanenko+etal2019c}. 
In the lossless case (with $A_{\tau l}=1$, $B_{\tau l}=-1/2$ and $C_{\tau l}=0$) we obtain the divergent series 
$Q_{\mrm{t},L}^\mrm{opt}=4L(L+2)/(k_0a)^2$, similar as in \eqref{eq:newboundQaLL}.

\section{Numerical examples}

As a numerical example we consider the atmospheric absorption by oxygen in the 60 GHz band \cite{Vleck1947,Meeks+Lilley1963,Tretyakov+etal2005},
which is of great importance due to its capability to mitigate interference in short range communication systems \cite{Park+Rappaport2007,Hawkins+etal1985,Wang+etal2012}.
As an additional interesting property of the lossy atmosphere, we demonstrate below that it also ultimately limits any super resolution capabilities of small antennas.

\begin{figure}[htb]
\begin{center}
\includegraphics[width=0.47\textwidth]{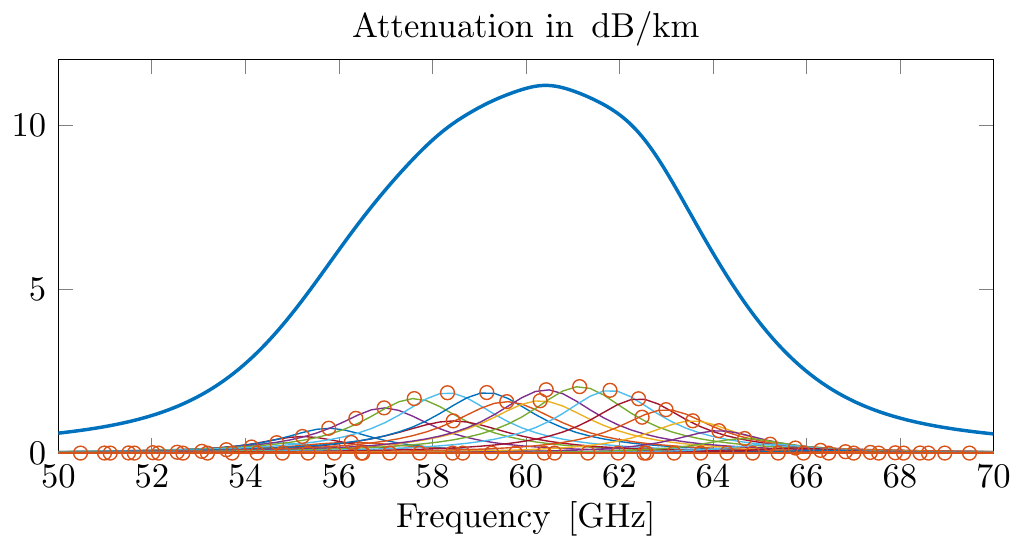}
\end{center}
\vspace{-5mm}
\caption{Absorption coefficient of dry air at normal pressure in the 60 GHz oxygen band (blue solid line). The solid lines indicated with an `o' show the pressure broadened
Lorentz profiles for each individual magnetic dipole transition of oxygen at normal pressure in air.
}
\label{fig:matfig31}
\end{figure}

In Fig.~\ref{fig:matfig31} is shown the absorption coefficient $\alpha=N\sigma_\mrm{a}$ of air (plotted in \unit{dB/km}) where $N$ is the number density of the absorptive gas (oxygen)
and $\sigma_\mrm{a}$ the absorption cross section per molecule.  The data is taken from the open source database HITRAN (high-resolution transmission molecular absorption database) \cite{Gordon+etal2017}. Due to the magnetic dipole moment of oxygen and its related rotational transitions, the background medium is magnetic
with $\alpha=2k_0\Im\{\sqrt{\mu}\}$ and permeability $\mu=1+\iu\mu^{\prime\prime}$ where $\mu^{\prime\prime}=\alpha/k_0$, assuming that $\mu^{\prime\prime}$ is small.
In the numerical examples below we have used $\mu^{\prime\prime}=2\cdot 10^{-6}$ corresponding approximately to the peak absorption seen in Fig.~\ref{fig:matfig31}.

\begin{figure}[htb]
\begin{center}
\includegraphics[width=0.48\textwidth]{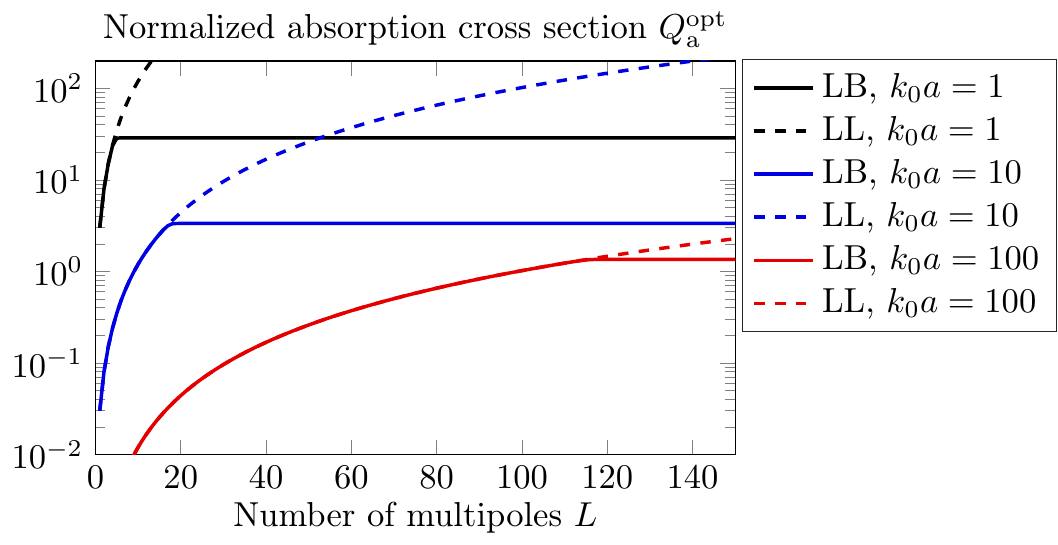}
\end{center}
\vspace{-5mm}
\caption{Optimal normalized absorption cross section $Q_\mrm{a}^\mrm{opt}$ as a function of maximal multipole order $L$, and a lossy background (LB) atmosphere
with $\mu^{\prime\prime}=2\cdot 10^{-6}$.
The dashed lines show the optimal absorption cross section for the lossless case (LL) $Q_{\mrm{a},L}^\mrm{opt}=L(L+2)/(k_0a)^2$.
}
\label{fig:matfig802}
\end{figure}

\begin{figure}[htb]
\begin{center}
\includegraphics[width=0.48\textwidth]{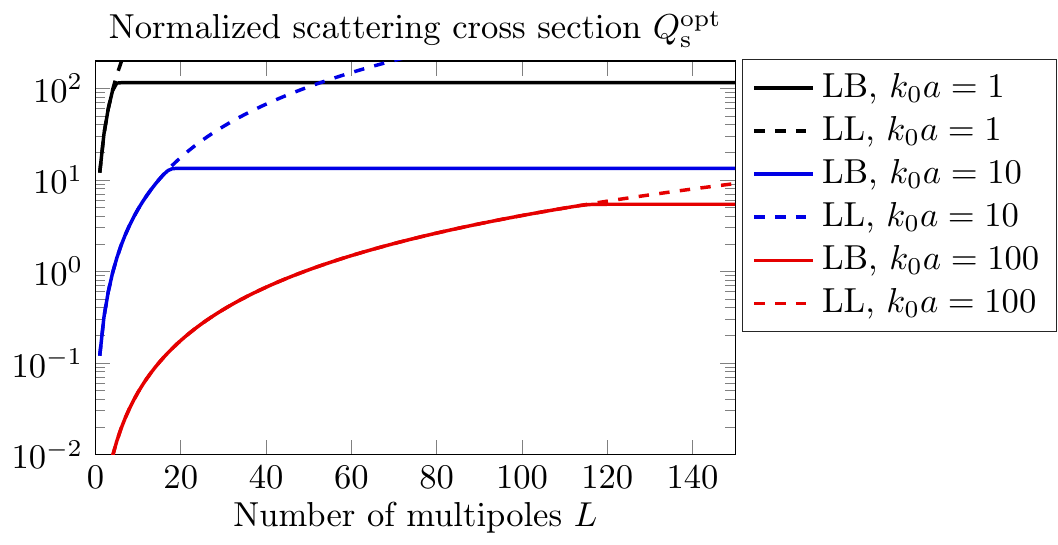}
\end{center}
\vspace{-5mm}
\caption{Optimal normalized scattering cross section $Q_\mrm{s}^\mrm{opt}$ as a function of maximal multipole order $L$, and a lossy background (LB) atmosphere
with $\mu^{\prime\prime}=2\cdot 10^{-6}$.
The dashed lines show the optimal scattering cross section for the lossless case (LL) $Q_{\mrm{s},L}^\mrm{opt}=4L(L+2)/(k_0a)^2$.
}
\label{fig:matfig803}
\end{figure}

\begin{figure}[ht]
\begin{center}
\includegraphics[width=0.49\textwidth]{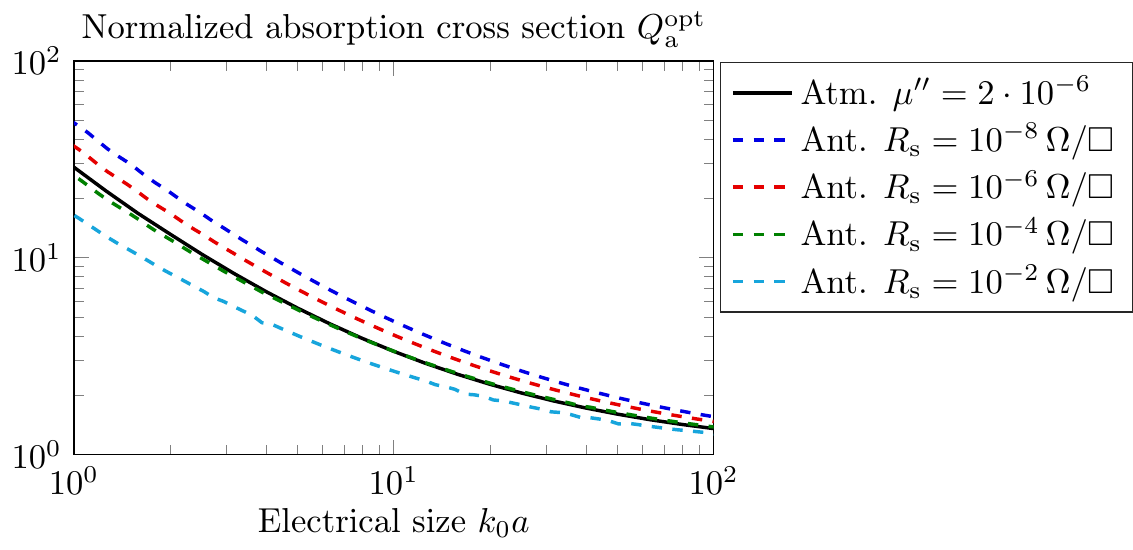}
\end{center}
\vspace{-5mm}
\caption{Comparison of the optimal absorption cross section $Q_\mrm{a}^\mrm{opt}$ \eqref{eq:newboundQa} for the lossy atmosphere (Atm.) with $\mu^{\prime\prime}=2\cdot 10^{-6}$,  
and the maximum effective area for an externally tuned spherical shell (Ant.) with surface resistivity $R_\mrm{s}=10^{-8},10^{-6},10^{-4},10^{-2}\Ohmps$ 
and a lossless atmosphere \cite{Gustafsson+Miloslav2019}. 
}
\label{fig:matfig804}
\end{figure}

In Figs.~\ref{fig:matfig802} and \ref{fig:matfig803} are shown the optimal normalized absorption and scattering cross sections $Q_\mrm{a}^\mrm{opt}$ 
and $Q_\mrm{s}^\mrm{opt}$ given by \eqref{eq:newboundQa} and \eqref{eq:newboundQs}, respectively. The graphs are calculated for 
a lossy background medium with $\epsilon=1$ and $\mu=1+\iu 2\cdot 10^{-6}$ (atmospheric oxygen),
the electrical sizes $k_0a=1,10,100$ (corresponding to $a=0.08,0.8,8 \unit{cm}$) and maximal multipole order $L=1,\ldots,150$ where $l\leq L$.
Finally, in Fig.~\ref{fig:matfig804} we illustrate that the constraints on absorption from the lossy atmosphere are comparable to those imposed on the antenna effective area 
by a surface resistivity of $R_{\mrm{s}}=10^{-4}\Ohmps$ on a spherical shell, \cf \cite[Fig.~3 on p.~5285]{Gustafsson+Miloslav2019}.
In practice, this means that it would be meaningless to design a ``super conducting'' antenna at 60\unit{GHz} with a surface resistivity
smaller than $R_{\mrm{s}}=10^{-4}\Ohmps$, see also \cite[Fig.~7 on p.~5287]{Gustafsson+Miloslav2019}.

\section{Conclusion}
In this contribution we have reformulated and extended some recent analytical results concerning 
an optical theorem and the associated optimal extinction, scattering and absorption in lossy surrounding media.
In essence, the theory gives new fundamental bounds that limits any single frequency super resolution capabilities
for a scatterer in a lossy surrounding media, with possible applications \eg
with antennas and resonators in communications, plasmonics  and medicin.





%


\begin{thebibliography}{10}
\providecommand{\url}[1]{#1}
\csname url@samestyle\endcsname
\providecommand{\newblock}{\relax}
\providecommand{\bibinfo}[2]{#2}
\providecommand{\BIBentrySTDinterwordspacing}{\spaceskip=0pt\relax}
\providecommand{\BIBentryALTinterwordstretchfactor}{4}
\providecommand{\BIBentryALTinterwordspacing}{\spaceskip=\fontdimen2\font plus
\BIBentryALTinterwordstretchfactor\fontdimen3\font minus
  \fontdimen4\font\relax}
\providecommand{\BIBforeignlanguage}[2]{{%
\expandafter\ifx\csname l@#1\endcsname\relax
\typeout{** WARNING: IEEEtran.bst: No hyphenation pattern has been}%
\typeout{** loaded for the language `#1'. Using the pattern for}%
\typeout{** the default language instead.}%
\else
\language=\csname l@#1\endcsname
\fi
#2}}
\providecommand{\BIBdecl}{\relax}
\BIBdecl

\bibitem{Bohren+Gilra1979}
C.~F. Bohren and D.~P. Gilra, ``Extinction by a spherical particle in an
  absorbing medium,'' \emph{J. Colloid Interface Sci.}, vol.~72, no.~2, pp.
  215--221, 1979.

\bibitem{Lebedev+etal1999}
A.~Lebedev, M.~Gartz, U.~Kreibig, and O.~Stenzel, ``Optical extinction by
  spherical particles in an absorbing medium: Application to composite
  absorbing films,'' \emph{Eur. Phys. J. D.}, vol.~6, no.~3, pp. 365--373,
  1999.

\bibitem{Sudiarta+Chylek2001}
I.~W. Sudiarta and P.~Chylek, ``Mie-scattering formalism for spherical
  particles embedded in an absorbing medium,'' \emph{J. Opt. Soc. Am. A},
  vol.~18, no.~6, pp. 1275--1278, 2001.

\bibitem{Durant+etal2007a}
S.~Durant, O.~Calvo-Perez, N.~Vukadinovic, and J.-J. Greffet, ``Light
  scattering by a random distribution of particles embedded in absorbing media:
  diagrammatic expansion of the extinction coefficient,'' \emph{J. Opt. Soc.
  Am. A}, vol.~24, no.~9, pp. 2943--2952, 2007.

\bibitem{Nordebo+etal2019a}
S.~Nordebo, G.~Kristensson, M.~Mirmoosa, and S.~Tretyakov, ``Optimal plasmonic
  multipole resonances of a sphere in lossy media,'' \emph{Phys. Rev. B},
  vol.~99, no.~5, p. 054301, 2019.

\bibitem{Nordebo+etal2019b}
S.~Nordebo, M.~Mirmoosa, and S.~Tretyakov, ``On the quasistatic optimal
  plasmonic resonances in lossy media,'' \emph{J. Appl. Phys.}, vol. 125, p.
  103105, 2019.

\bibitem{Ivanenko+etal2019c}
Y.~Ivanenko, M.~Gustafsson, and S.~Nordebo., ``Optical theorems and physical
  bounds on absorption in lossy media,'' \emph{Opt. Express}, vol.~27, no.~23,
  pp. 34\,323--34\,342, 2019.

\bibitem{Park+Rappaport2007}
C.~Park and T.~S. Rappaport, ``Short-range wireless communications for
  next-generation networks: {UWB}, 60 {GHz} millimeter-wave {WPAN}, and
  {ZigBee},'' \emph{IEEE Wireless Communications}, pp. 70--78, 2007.

\bibitem{Hawkins+etal1985}
N.~D. Hawkins, R.~Steele, D.~C. Rickard, and C.~R. Shepherd, ``Path loss
  characteristics of 60 {GHz} transmissions,'' \emph{Elect. Lett.}, vol.~21,
  no.~22, pp. 1054--1055, 1985.

\bibitem{Wang+etal2012}
J.~Wang, H.~Zhang, T.~Lv, and T.~A. Gulliver, ``Capacity of 60 {GHz} wireless
  communication systems over fading channels,'' \emph{Journal of networks},
  vol.~7, no.~1, pp. 203--209, 2012.

\bibitem{Karlsson2004}
A.~Karlsson, ``Physical limitations of antennas in a lossy medium,'' \emph{IEEE
  Trans. Antennas Propagat.}, vol.~52, pp. 2027--2033, 2004.

\bibitem{Skrivervik2019}
A.~K. Skrivervik, M.~Bosiljevac, and Z.~Sipus, ``Fundamental limits for
  implanted antennas: Maximum power density reaching free space,'' \emph{IEEE
  Trans. Antennas Propagat.}, vol.~67, no.~8, pp. 4978--4988, 2019.

\bibitem{Gustafsson+Miloslav2019}
M.~Gustafsson and M.~Capek, ``Maximum gain, effective area, and directivity,''
  \emph{IEEE Trans. Antennas Propagat.}, vol.~67, no.~8, pp. 5282--5293, 2019.

\bibitem{Maier2007}
S.~A. Maier, \emph{Plasmonics: Fundamentals and Applications}.\hskip 1em plus
  0.5em minus 0.4em\relax Berlin: Sprin\-ger-Verlag, 2007.

\bibitem{Maslovski+etal2016}
S.~I. Maslovski, C.~R. Simovski, and S.~A. Tretyakov, ``Overcoming black body
  radiation limit in free space: metamaterial superemitter,'' \emph{New J.
  Phys.}, vol.~18, p. 013034, 2016.

\bibitem{Valagiannopoulos+Tretyakov2016}
C.~A. Valagiannopoulos and S.~A. Tretyakov, ``Theoretical concepts of
  unlimited-power reflectors, absorbers, and emitters with conjugately matched
  layers,'' \emph{Phys. Rev. B}, vol.~94, p. 125117, 2016.

\bibitem{Huang+etal2008}
X.~Huang, P.~K. Jain, I.~H. El-Sayed, and M.~A. El-Sayed, ``Plasmonic
  photothermal therapy ({PPTT}) using gold nanoparticles,'' \emph{Lasers Med
  Sci}, vol.~23, pp. 217--228, 2008.

\bibitem{Sassaroli+etal2012}
E.~Sassaroli, K.~C.~P. Li, and B.~E. O'Neil, ``Radio frequency absorption in
  gold nanoparticle suspensions: a phenomenological study,'' \emph{J. Phys. D:
  Appl. Phys.}, vol.~45, pp. 1--15, 2012, 075303.

\bibitem{Nordebo+etal2017a}
S.~Nordebo, M.~Dalarsson, Y.~Ivanenko, D.~Sj\"{o}berg, and R.~Bayford, ``On the
  physical limitations for radio frequency absorption in gold nanoparticle
  suspensions,'' \emph{J. Phys. D: Appl. Phys.}, vol.~50, no.~15, p. 155401,
  2017.

\bibitem{Jackson1999}
J.~D. Jackson, \emph{Classical Electrodynamics}, 3rd~ed.\hskip 1em plus 0.5em
  minus 0.4em\relax New York: John Wiley \& Sons, 1999.

\bibitem{Kristensson2016}
G.~Kristensson, \emph{Scattering of Electromagnetic Waves by Obstacles}.\hskip
  1em plus 0.5em minus 0.4em\relax SciTech Publishing, Edison, NJ, 2016.

\bibitem{Bohren+Huffman1983}
C.~F. Bohren and D.~R. Huffman, \emph{Absorption and Scattering of Light by
  Small Particles}.\hskip 1em plus 0.5em minus 0.4em\relax New York: John Wiley
  \& Sons, 1983.

\bibitem{Vleck1947}
J.~H.~V. Vleck, ``The absorption of microwaves by oxygen,'' \emph{Phys. Rev.},
  vol.~71, no.~7, pp. 413--424, 1947.

\bibitem{Meeks+Lilley1963}
M.~L. Meeks and A.~E. Lilley, ``The microwave spectrum of oxygen in the
  {E}arth's atmosphere,'' \emph{Journal of Geophysical Research}, vol.~68,
  no.~6, pp. 1683--1703, 1963.

\bibitem{Tretyakov+etal2005}
M.~Y. Tretyakov, M.~A. Koshelev, V.~V. Dorovskikh, D.~S. Makarov, and P.~W.
  Rosenkranz, ``60-{GHz} oxygen band: precise broadening and central
  frequencies of fine-structure lines, absolute absorption profile at
  atmospheric pressure, and revision of mixing coefficients,'' \emph{Journal of
  Molecular Spectroscopy}, vol. 231, pp. 1--14, 2005.

\bibitem{Gordon+etal2017}
I.~E. Gordon and et~al, ``The {HITRAN2016} molecular spectroscopic database,''
  \emph{Journal of Quantitative Spectroscopy \& Radiative Transfer}, vol. 203,
  pp. 3--69, 2017.

\end{thebibliography}

\end{document}